\documentclass[aip,apl,reprint]{revtex4-1}

\usepackage{amsmath}
\usepackage{amsfonts}
\usepackage{amssymb}
\usepackage{graphicx}
\usepackage{graphics}
\usepackage{hyperref}
\begin{document}

%\preprint{}

\title{Surface-electrode ion trap with integrated light source} %Title of paper



\author{Tony Hyun Kim}
\email{kimt@mit.edu}
\author{Peter F. Herskind}
\author{Isaac L. Chuang}
\affiliation{Center for Ultracold Atoms, Department of Physics, Massachusetts Institute of Technology\\
77 Massachusetts Avenue, Cambridge, MA 02139}


\date{\today}

\begin{abstract}
An atomic ion is trapped at the tip of a single-mode optical fiber in a cryogenic ($8$~K) surface-electrode ion trap. The fiber serves as an integrated source of laser light, which drives the quadrupole qubit transition of $^{88}$Sr$^+$. Through \emph{in situ} translation of the nodal point of the trapping field, the Gaussian beam profile of the fiber output is imaged, and the fiber-ion displacement, in units of the mode waist at the ion, is optimized to within $0.13\pm0.10$ of the mode center despite an initial offset of $3.30\pm0.10$. Fiber-induced charging at $125~\mu$W is observed to be ${\sim}10$~V/m at an ion height of $670~\mu$m, with charging and discharging time constants of $1.6\pm0.3$~s and $4.7\pm0.6$~s respectively. This work is of importance to large-scale, ion-based quantum information processing, where optics integration in surface-electrode designs may be a crucial enabling technology. 
\end{abstract}


\maketitle %\maketitle must follow title, authors, abstract and \pacs

%%%%%%%%%%%%%%%%%%%%%%%%%%%%%%%%%%%%%%%%%%%%%%%%%%%%%%%%%%%%%%%%%%%%%%%%%%%%%
% Introduction and motivation
An array of trapped ions in optical cavities, connected by a network of optical fibers, represents a possible distributed architecture for large-scale quantum information processing~\cite{Cirac1997} (QIP). Due to the necessity of efficient light collection, laser cooling and qubit state manipulation, the realization of a quantum network or processor at the level of tens and hundreds of qubits strongly motivates the integration of optics in surface-electrode ion traps~\cite{Kim2005}. However, the potential benefits of integrated optics have long been overshadowed by the challenge of trapping ions in the proximity of dielectrics~\cite{Harlander2010}, as well as the difficulty of guaranteeing good spatial overlap of the trapped ion with the field mode of the integrated element.

In the past, there have been demonstrations of integration of bulk mirrors~\cite{Guthohrlein2001,Herskind2009a,Shu2010}, multi-mode (MM) optical fibers~\cite{Wilson2011}, and phase-Fresnel lenses~\cite{Streed2011} into radio frequency (RF) traps with three-dimensional electrodes. More recently, integration of MM fibers~\cite{VanDevender2010} and microscopic reflective optics~\cite{Herskind2010} for collection of ion fluorescence has been demonstrated in microfabricated surface-electrode traps. Complementing such efforts on light collection, the present work demonstrates light delivery through an integrated single-mode (SM) fiber in a scalable, surface-electrode design, and an \emph{in situ} micrometer-scale positioning of the ion relative to the integrated structure. Future developments in optics integration, such as microcavities for the realization of quantum light-matter interfaces~\cite{Cirac1997,Kim2009}, or lensed fibers for faster gate times and optical trapping of ions~\cite{SchneiderCh2010}, will employ sub-$10~\mu$m waists~\cite{Herskind2010}, underscoring the importance of \emph{in situ} ion positioning~\cite{Herskind2009}.

We report on the construction of a fiber-trap system, and demonstrate the ability of the integrated light source to drive the $674$~nm quadrupole transition of $^{88}$Sr$^+$. The quadrupole transition is of particular interest in QIP with trapped ions, where it serves as the optical qubit~\cite{Blatt2008}, as well as in metrology, where it constitutes an optical frequency standard~\cite{Margolis2004}. The ion-fiber spatial overlap is optimized \emph{in situ} by micromotion-free translation of the ion using segmented RF electrodes. We use this technique to map out the Gaussian profile of the fiber mode along a single transverse axis. With the ion positioned over the peak of the mode, we quantify the magnitude and timescale of fiber-induced charging.

%%%%%%%%%%%%%%%%%%%%%%%%%%%%%%%%%%%%%%%%%%%%%%%%%%%%%%%%%%%%%%%%%%%%%%%%%%%%%
% Trap construction
\begin{figure}[t]
\includegraphics[width=1\columnwidth]{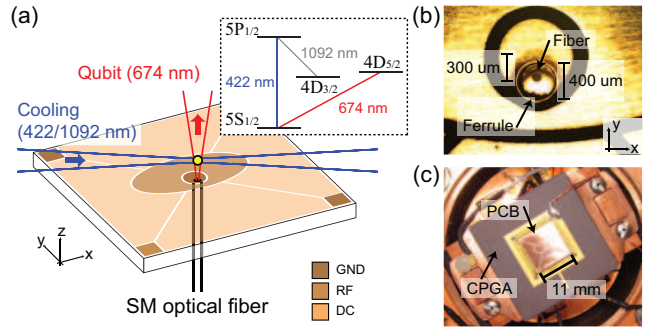}
\caption{(Color online) (a) Schematic of the surface-electrode ion trap with integrated optical fiber. The $^{88}$Sr$^+$ qubit laser is delivered axially (along $z$) through the fiber, while Doppler cooling beams propagate in the horizontal $xy$-plane. (b) Alignment of the optical ferrule with respect to the trap electrodes. The ferrule is rotated until the fiber is aligned with the minor ($y$) axis of the trap. (c) The fiber-trap system mounted on a CPGA and installed on the $8$ K basplate of a closed-cycle cryostat.\label{fig:trap}}
\end{figure}

Fiber-trap integration is achieved by embedding the fiber within the trap substrate. Fig. \ref{fig:trap}(a) shows a schematic of the ion trap design, which is a modified version of the surface-electrode point Paul trap described recently~\cite{KimH2010}. The center, grounded electrode has a diameter of $1.1$~mm. The elliptical RF pad has major- and minor-axis diameters of $5.9$ mm and $2.8$ mm, respectively, and is shifted by $500~\mu$m along the minor-axis relative to the center of the ground electrode. Electrode gaps are $100~\mu$m. This design achieves an ion height of $670~\mu$m and the electrode asymmetries uniquely define the principal axes of the trap, which are tilted by $30^\circ$ in the $yz$-plane for efficient cooling and micromotion compensation. The side electrodes are used for DC compensation of stray electric fields, as well as radial translation of the RF node by use of additional RF voltages. The trap is defined on a printed circuit board (PCB) with copper electrodes on a low-RF-loss substrate (Rogers 4350B, fabricated by Hughes circuits). The PCB includes a $400~\mu$m-diameter plated via in the center ground electrode for the insertion of an optical ferrule. The via is offset by $300~\mu$m with respect to the ground electrode to account for the displacement of the trapping point that accompanies the shift of the elliptical RF electrode. 

The optical fiber (OZ Optics, PMF-633-4/125-3-L) is SM for $674$~nm and is conventionally prepared (i.e. cured in fiber epoxy and polished) in a stainless steel SMA ferrule whose tip has been machined to match the $400~\mu$m-diameter of the PCB via. The assembly of PCB and ferrule is performed under microscope, as in Fig.~\ref{fig:trap}(b), where machining imprecision of the ferrule is evident in the form of ${\sim}100~\mu$m nonconcentricity between the fiber and the ferrule. The ferrule is rotated with respect to the PCB to place the fiber roughly along the minor axis of the trap, and is cured using cyanoacrylate adhesive.

%%%%%%%%%%%%%%%%%%%%%%%%%%%%%%%%%%%%%%%%%%%%%%%%%%%%%%%%%%%%%%%%%%%%%%%%%%%%%
% Experimental setup
The fiber-trap system is installed on a ceramic pin grid array (CPGA) and mounted on the $8$~K baseplate of a closed-cycle cryostat~\cite{Antohi2009} as shown in Fig. \ref{fig:trap}(c). The fiber is routed through a hole in the CPGA and a hole in a flange of the vacuum chamber, where it is sealed in place with TorrSeal UHV epoxy. The trap is operated at a typical RF frequency of $2\pi\times 6$~MHz and $250$~Vpp amplitude, achieving secular frequencies of $\omega_{z'}=2\pi\times 410$~kHz, $\omega_x = 2\pi\times 240$~kHz, $\omega_{y'} = 2\pi\times 170$~kHz. We produce $^{88}$Sr$^+$ ions by resonant photoionization~\cite{Brownnutt2007}, which are Doppler cooled on the $5$S$_{1/2}$$\leftrightarrow$5P$_{1/2}$ transition at $422$~nm, while simultaneously driving the 4D$_{3/2}$$\leftrightarrow$5P$_{1/2}$ transition at $1092$~nm. Ion fluorescence at $422$~nm is collected by a $0.5$~NA lens inside the chamber and imaged onto a CCD camera and a photomultiplier tube (PMT), both with individual ion resolution.
\begin{figure}[t]
\includegraphics[width=1\columnwidth]{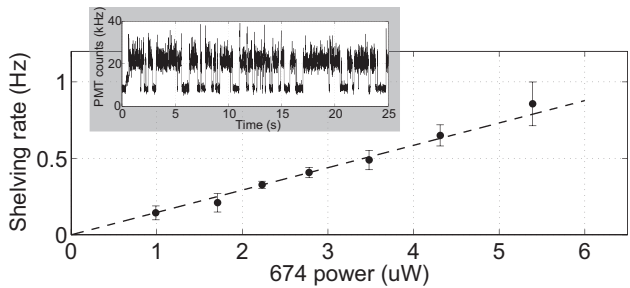}
\caption{Shelving rate as a function of $674$~nm power coupled to the trap fiber. Inset: telegraph log of a single trapped ion as it is shelved into the dark $4$D$_{5/2}$ state by the fiber light.\label{fig:telegraph}}
\end{figure}

%%%%%%%%%%%%%%%%%%%%%%%%%%%%%%%%%%%%%%%%%%%%%%%%%%%%%%%%%%%%%%%%%%%%%%%%%%%%%
% 674 experiment and results
Interaction between the ion and the fiber mode is demonstrated using the electron shelving method~\cite{Dehmelt1975}. The ion is driven on the 5S$_{1/2}$$\leftrightarrow$4D$_{5/2}$ transition by $674$~nm light from the fiber while being simultaneously illuminated by the $422$~nm and $1092$~nm beams. Upon shelving to the 4D$_{5/2}$ state, no $422$~nm photons are scattered and the ion remains dark until it decays spontaneously back to the 5S$_{1/2}$ state, as illustrated by the single-ion telegraph in the inset of Fig. \ref{fig:telegraph}. An effective shelving rate can be quantified by counting the bright-to-dark transitions per total bright time. In Fig.~\ref{fig:telegraph}, the linear relationship between shelving rate and $674~$nm power coupled to the fiber is shown, as expected for a weak driving field.

Because of the exponential fall-off in intensity along the transverse plane of a Gaussian mode, a method for \emph{in situ} control of the ion positioning is highly desired. While DC potentials may achieve ion translation, the resultant displacement of the ion from the RF node incurs additional micromotion that broadens atomic transitions~\cite{Berkeland1998}, which significantly limits the range and usefulness of DC translation. In contrast, micromotion-free translation can be achieved by shifting the quadrupole field node itself, as has been demonstrated recently~\cite{Herskind2009,VanDevender2010,KimH2010}, by using multiple RF-voltages applied to different electrodes of the trap.

\begin{figure}[t]
\includegraphics[width=1\columnwidth]{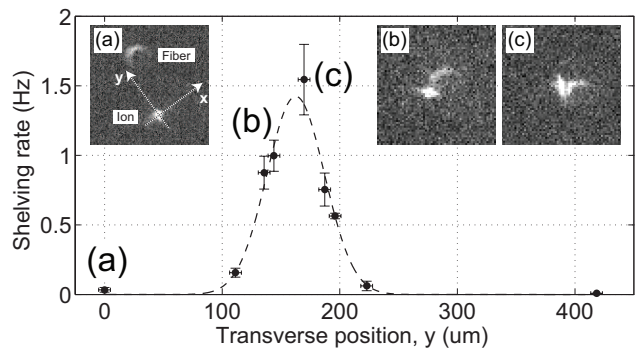}
\caption{Measurement of the mode profile of the integrated fiber using the ion as a probe. The dashed line is a fit to a Gaussian profile centered at $163.4\pm0.8~\mu$m, where the transverse axis has been calibrated by an independent measurement of the fiber mode size at the ion height. Images show the relative position of a single ion and an unfocused image of the fiber.\label{fig:modemeas}}
\end{figure}

We utilize multiple RF sources to achieve micromotion-free translation of the ion in the horizontal plane of a surface-electrode ion trap. Fig.~\ref{fig:modemeas} shows the change in shelving rate as the ion is translated along the $y$-axis, across the mode of the fiber. The dashed line represents a Gaussian beam shape indicating good qualitative agreement with the shelving rate profile. The beam waist at the ion height of $670~\mu$m has been measured independently using an identical fiber to be $50~\mu$m, which is used to calibrate the $y$-axis in Fig.~\ref{fig:modemeas}. In units of the measured mode waist, the ion is brought to within $0.13\pm0.10$ of the mode center, despite an initial displacement of $3.34\pm0.10$ arising from trap construction. CCD images show the ion displaced relative to the (unfocused) image of the fiber. In all measurements, the ion was positioned at the RF node by eliminating micromotion amplitude according to the correlation measurement technique~\cite{Berkeland1998}.

\begin{figure}[t]
\includegraphics[width=1\columnwidth]{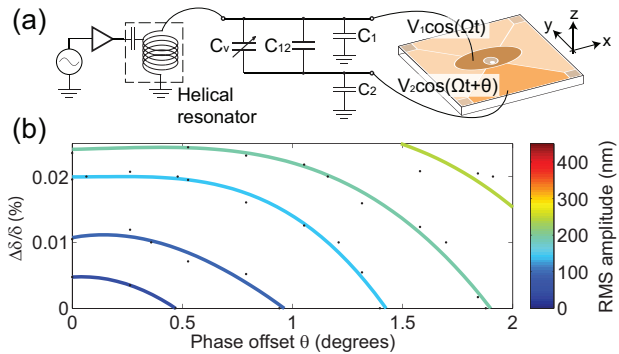}
\caption{(Color online) (a) A circuit model for implementation of two in-phase RF drives through a capacitive network. Variable capacitor $C_v$ is used to adjust the RF ratio $\delta=V_2/V_1$. (b) Results of a Monte Carlo simulation showing the RMS amplitudes of classical ion trajectories under nonideal RF control $(\Delta\delta/\delta, \theta)$ at the operational point (c) of Fig. \ref{fig:modemeas}. Contour lines indicate steps of $50~$nm.\label{fig:phasediag}}
\end{figure}

The ultimate precision of RF translation is limited by the stability and control of the relative amplitudes and phases between the multiple sources. The position uncertainty indicated in Fig. \ref{fig:modemeas} is limited by the resolution of our imaging system to $\pm 5~\mu$m and is not fundamental to the RF node translation method. Fig. \ref{fig:phasediag}(a) shows our implementation for achieving two in-phase RF voltages through a passive network. Capacitances $C_1, C_2 \approx 30~$pF are intrinsic to the trap electrodes and cryostat wiring, as is $C_{12} \approx 3~$pF, which accounts for the intrinsic capacitive coupling between the two electrodes. We introduce a mechanically tunable capacitor $C_v=0.5-30$~pF (Voltronics Corp.) in order to adjust the RF ratio $\delta=V_2/V_1$. Fig. \ref{fig:phasediag}(b) is the results of a Monte Carlo simulation at the operational point (c) of Fig. \ref{fig:modemeas}, showing RMS amplitudes of classical ion trajectories in steps of $50~$nm under RF ratio ($\Delta\delta/\delta$) and phase ($\theta$) imperfections. Given a typical temperature coefficient of $50$~ppm$/$C$^\circ$ for $C_v$ we expect a ratio imprecision of $\Delta\delta/\delta = 0.005\%/$C$^\circ$, while phase error $\theta$ arises from differential resistances of the two wire paths at a negligible $0.06^\circ$ per differential ohm. With a $1^\circ$C control in capacitor temperature, we conclude that the current implementation achieves ${\sim}50~$nm RMS positioning precision. Furthermore, the sensitivity to RF imperfections is trap-design dependent, and may be reduced at the cost of total translational range.

With the ion centered in the fiber mode, we have looked for effects of dielectric charging by the $674$~nm fiber light~\cite{Harlander2010}. In these tests, $125~\mu$W of $674$~nm light is coupled into the fiber while the amplitude of ion micromotion is recorded for several minutes to detect any dynamic shifts in the ion position due to a possible fiber-induced generation of charge. We observe induced fields of ${\sim}10$~V/m by the fiber, with charging and discharging time constants of $1.6\pm0.3$~s and $4.7\pm0.6$~s. Following the initial charge generation, the micromotion amplitude remains constant for minutes, indicating stable saturation of fiber-induced charge.

%%%%%%%%%%%%%%%%%%%%%%%%%%%%%%%%%%%%%%%%%%%%%%%%%%%%%%%%%%%%%%%%%%%%%%%%%%%%%
% Conclusions
In conclusion, we have demonstrated an ion trap with an integrated SM fiber for light delivery, and an \emph{in situ} micromotion-free optimization of the ion-fiber spatial overlap. The fiber has been used to directly address the qubit transition of a single ion in the trap and, as such, the fiber-trap system represents a step towards optics-integration for large-scale QIP in surface-electrode designs. Moreover, our assembly is compatible with more advanced fiber systems, such as lensed fibers to achieve higher field intensities, or photonic crystal fibers that propagate all relevant lasers through a single integrated port. In a cryogenic environment, such an all-inclusive port may eliminate the requirement of free-space optical access, greatly reducing the heatload and enabling sub-Kelvin trap operation. An integrated trap that utilizes the fiber facet as one of the mirrors of an optical cavity could serve as a node in a distributed QIP architecture where the photon state can be extracted through the fiber~\cite{Cirac1997,Kim2009}.

\begin{acknowledgments}  
T.H.K. was supported by the Siebel and Chorafas Foundations. P.F.H. is grateful for the support from the Carlsberg and Lundbeck Foundations.
\end{acknowledgments}

%\bibliography{bibliography}

\begin{thebibliography}{20}%
\makeatletter
\providecommand \@ifxundefined [1]{%
 \@ifx{#1\undefined}
}%
\providecommand \@ifnum [1]{%
 \ifnum #1\expandafter \@firstoftwo
 \else \expandafter \@secondoftwo
 \fi
}%
\providecommand \@ifx [1]{%
 \ifx #1\expandafter \@firstoftwo
 \else \expandafter \@secondoftwo
 \fi
}%
\providecommand \natexlab [1]{#1}%
\providecommand \enquote  [1]{``#1''}%
\providecommand \bibnamefont  [1]{#1}%
\providecommand \bibfnamefont [1]{#1}%
\providecommand \citenamefont [1]{#1}%
\providecommand \href@noop [0]{\@secondoftwo}%
\providecommand \href [0]{\begingroup \@sanitize@url \@href}%
\providecommand \@href[1]{\@@startlink{#1}\@@href}%
\providecommand \@@href[1]{\endgroup#1\@@endlink}%
\providecommand \@sanitize@url [0]{\catcode `\\12\catcode `\$12\catcode
  `\&12\catcode `\#12\catcode `\^12\catcode `\_12\catcode `\%12\relax}%
\providecommand \@@startlink[1]{}%
\providecommand \@@endlink[0]{}%
\providecommand \url  [0]{\begingroup\@sanitize@url \@url }%
\providecommand \@url [1]{\endgroup\@href {#1}{\urlprefix }}%
\providecommand \urlprefix  [0]{URL }%
\providecommand \Eprint [0]{\href }%
\providecommand \doibase [0]{http://dx.doi.org/}%
\providecommand \selectlanguage [0]{\@gobble}%
\providecommand \bibinfo  [0]{\@secondoftwo}%
\providecommand \bibfield  [0]{\@secondoftwo}%
\providecommand \translation [1]{[#1]}%
\providecommand \BibitemOpen [0]{}%
\providecommand \bibitemStop [0]{}%
\providecommand \bibitemNoStop [0]{.\EOS\space}%
\providecommand \EOS [0]{\spacefactor3000\relax}%
\providecommand \BibitemShut  [1]{\csname bibitem#1\endcsname}%
\let\auto@bib@innerbib\@empty
%</preamble>
\bibitem [{\citenamefont {Cirac}\ \emph {et~al.}(1997)\citenamefont {Cirac},
  \citenamefont {Zoller}, \citenamefont {Kimble},\ and\ \citenamefont
  {Mabuchi}}]{Cirac1997}%
  \BibitemOpen
  \bibfield  {author} {\bibinfo {author} {\bibfnamefont {J.~I.}\ \bibnamefont
  {Cirac}}, \bibinfo {author} {\bibfnamefont {P.}~\bibnamefont {Zoller}},
  \bibinfo {author} {\bibfnamefont {H.~J.}\ \bibnamefont {Kimble}}, \ and\
  \bibinfo {author} {\bibfnamefont {H.}~\bibnamefont {Mabuchi}},\ }\href
  {http://prola.aps.org/abstract/PRL/v78/i16/p3221_1} {\bibfield  {journal}
  {\bibinfo  {journal} {Physical Review Letters}\ }\textbf {\bibinfo {volume}
  {78}},\ \bibinfo {pages} {3221} (\bibinfo {year} {1997})}\BibitemShut
  {NoStop}%
\bibitem [{\citenamefont {Kim}\ \emph {et~al.}(2005)\citenamefont {Kim},
  \citenamefont {Pau}, \citenamefont {Ma}, \citenamefont {McLellan},
  \citenamefont {Gates}, \citenamefont {Kornblit}, \citenamefont {Slusher},
  \citenamefont {Jopson}, \citenamefont {Kang},\ and\ \citenamefont
  {Dinu}}]{Kim2005}%
  \BibitemOpen
  \bibfield  {author} {\bibinfo {author} {\bibfnamefont {J.}~\bibnamefont
  {Kim}}, \bibinfo {author} {\bibfnamefont {S.}~\bibnamefont {Pau}}, \bibinfo
  {author} {\bibfnamefont {Z.}~\bibnamefont {Ma}}, \bibinfo {author}
  {\bibfnamefont {H.~R.}\ \bibnamefont {McLellan}}, \bibinfo {author}
  {\bibfnamefont {J.~V.}\ \bibnamefont {Gates}}, \bibinfo {author}
  {\bibfnamefont {A.}~\bibnamefont {Kornblit}}, \bibinfo {author}
  {\bibfnamefont {R.~E.}\ \bibnamefont {Slusher}}, \bibinfo {author}
  {\bibfnamefont {R.~M.}\ \bibnamefont {Jopson}}, \bibinfo {author}
  {\bibfnamefont {I.}~\bibnamefont {Kang}}, \ and\ \bibinfo {author}
  {\bibfnamefont {M.}~\bibnamefont {Dinu}},\ }\href
  {http://www.rinton.net/xqic5/qic-5-7/515-537.pdf} {\bibfield  {journal}
  {\bibinfo  {journal} {Quantum Information \& Computation}\ }\textbf {\bibinfo
  {volume} {5}},\ \bibinfo {pages} {515} (\bibinfo {year} {2005})}\BibitemShut
  {NoStop}%
\bibitem [{\citenamefont {Harlander}\ \emph {et~al.}(2010)\citenamefont
  {Harlander}, \citenamefont {Brownnutt}, \citenamefont {Hansel},\ and\
  \citenamefont {Blatt}}]{Harlander2010}%
  \BibitemOpen
  \bibfield  {author} {\bibinfo {author} {\bibfnamefont {M.}~\bibnamefont
  {Harlander}}, \bibinfo {author} {\bibfnamefont {M.}~\bibnamefont
  {Brownnutt}}, \bibinfo {author} {\bibfnamefont {W.}~\bibnamefont {Hansel}}, \
  and\ \bibinfo {author} {\bibfnamefont {R.}~\bibnamefont {Blatt}},\
  }\href@noop {} {\bibfield  {journal} {\bibinfo  {journal} {New Journal of
  Physics}\ }\textbf {\bibinfo {volume} {12}},\ \bibinfo {pages} {093035}
  (\bibinfo {year} {2010})}\BibitemShut {NoStop}%
\bibitem [{\citenamefont {Guthohrlein}\ \emph {et~al.}(2001)\citenamefont
  {Guthohrlein}, \citenamefont {Keller}, \citenamefont {Hayasaka},
  \citenamefont {Lange},\ and\ \citenamefont {Walther}}]{Guthohrlein2001}%
  \BibitemOpen
  \bibfield  {author} {\bibinfo {author} {\bibfnamefont {G.~R.}\ \bibnamefont
  {Guthohrlein}}, \bibinfo {author} {\bibfnamefont {M.}~\bibnamefont {Keller}},
  \bibinfo {author} {\bibfnamefont {K.}~\bibnamefont {Hayasaka}}, \bibinfo
  {author} {\bibfnamefont {W.}~\bibnamefont {Lange}}, \ and\ \bibinfo {author}
  {\bibfnamefont {H.}~\bibnamefont {Walther}},\ }\href
  {http://www.nature.com/nature/journal/v414/n6859/abs/414049a0.html}
  {\bibfield  {journal} {\bibinfo  {journal} {Nature}\ }\textbf {\bibinfo
  {volume} {414}},\ \bibinfo {pages} {49} (\bibinfo {year} {2001})}\BibitemShut
  {NoStop}%
\bibitem [{\citenamefont {Herskind}\ \emph
  {et~al.}(2009{\natexlab{a}})\citenamefont {Herskind}, \citenamefont {Dantan},
  \citenamefont {Marler}, \citenamefont {Albert},\ and\ \citenamefont
  {Drewsen}}]{Herskind2009a}%
  \BibitemOpen
  \bibfield  {author} {\bibinfo {author} {\bibfnamefont {P.~F.}\ \bibnamefont
  {Herskind}}, \bibinfo {author} {\bibfnamefont {A.}~\bibnamefont {Dantan}},
  \bibinfo {author} {\bibfnamefont {J.~P.}\ \bibnamefont {Marler}}, \bibinfo
  {author} {\bibfnamefont {M.}~\bibnamefont {Albert}}, \ and\ \bibinfo {author}
  {\bibfnamefont {M.}~\bibnamefont {Drewsen}},\ }\href
  {http://www.nature.com/nphys/journal/v5/n7/full/nphys1302.html} {\bibfield
  {journal} {\bibinfo  {journal} {Nature Physics}\ }\textbf {\bibinfo {volume}
  {5}},\ \bibinfo {pages} {494} (\bibinfo {year}
  {2009}{\natexlab{a}})}\BibitemShut {NoStop}%
\bibitem [{\citenamefont {Shu}\ \emph {et~al.}(2010)\citenamefont {Shu},
  \citenamefont {Kurz}, \citenamefont {Dietrich},\ and\ \citenamefont
  {Blinov}}]{Shu2010}%
  \BibitemOpen
  \bibfield  {author} {\bibinfo {author} {\bibfnamefont {G.}~\bibnamefont
  {Shu}}, \bibinfo {author} {\bibfnamefont {N.}~\bibnamefont {Kurz}}, \bibinfo
  {author} {\bibfnamefont {M.}~\bibnamefont {Dietrich}}, \ and\ \bibinfo
  {author} {\bibfnamefont {B.~B.}\ \bibnamefont {Blinov}},\ }\href@noop {}
  {\bibfield  {journal} {\bibinfo  {journal} {Phys. Rev. A}\ }\textbf {\bibinfo
  {volume} {81}},\ \bibinfo {pages} {042321} (\bibinfo {year}
  {2010})}\BibitemShut {NoStop}%
\bibitem [{\citenamefont {Wilson}\ \emph {et~al.}(2011)\citenamefont {Wilson},
  \citenamefont {Takahashi}, \citenamefont {Riley-Watson}, \citenamefont
  {Orucevic}, \citenamefont {Blythe}, \citenamefont {Mortensen}, \citenamefont
  {Crick}, \citenamefont {Seymour-Smith}, \citenamefont {Brama}, \citenamefont
  {Keller},\ and\ \citenamefont {Lange}}]{Wilson2011}%
  \BibitemOpen
  \bibfield  {author} {\bibinfo {author} {\bibfnamefont {A.}~\bibnamefont
  {Wilson}}, \bibinfo {author} {\bibfnamefont {H.}~\bibnamefont {Takahashi}},
  \bibinfo {author} {\bibfnamefont {A.}~\bibnamefont {Riley-Watson}}, \bibinfo
  {author} {\bibfnamefont {F.}~\bibnamefont {Orucevic}}, \bibinfo {author}
  {\bibfnamefont {P.}~\bibnamefont {Blythe}}, \bibinfo {author} {\bibfnamefont
  {A.}~\bibnamefont {Mortensen}}, \bibinfo {author} {\bibfnamefont {D.~R.}\
  \bibnamefont {Crick}}, \bibinfo {author} {\bibfnamefont {N.}~\bibnamefont
  {Seymour-Smith}}, \bibinfo {author} {\bibfnamefont {E.}~\bibnamefont
  {Brama}}, \bibinfo {author} {\bibfnamefont {M.}~\bibnamefont {Keller}}, \
  and\ \bibinfo {author} {\bibfnamefont {W.}~\bibnamefont {Lange}},\
  }\href@noop {} {\bibfield  {journal} {\bibinfo  {journal} {arXiv:1101.5877}\
  } (\bibinfo {year} {2011})}\BibitemShut {NoStop}%
\bibitem [{\citenamefont {Streed}\ \emph {et~al.}(2011)\citenamefont {Streed},
  \citenamefont {Norton}, \citenamefont {Jechow}, \citenamefont {Weinhold},\
  and\ \citenamefont {Kielpinski}}]{Streed2011}%
  \BibitemOpen
  \bibfield  {author} {\bibinfo {author} {\bibfnamefont {E.~W.}\ \bibnamefont
  {Streed}}, \bibinfo {author} {\bibfnamefont {B.}~\bibnamefont {Norton}},
  \bibinfo {author} {\bibfnamefont {A.}~\bibnamefont {Jechow}}, \bibinfo
  {author} {\bibfnamefont {T.~J.}\ \bibnamefont {Weinhold}}, \ and\ \bibinfo
  {author} {\bibfnamefont {D.}~\bibnamefont {Kielpinski}},\ }\href@noop {}
  {\bibfield  {journal} {\bibinfo  {journal} {Phys. Rev. Lett.}\ }\textbf
  {\bibinfo {volume} {106}},\ \bibinfo {pages} {010502} (\bibinfo {year}
  {2011})}\BibitemShut {NoStop}%
\bibitem [{\citenamefont {VanDevender}\ \emph {et~al.}(2010)\citenamefont
  {VanDevender}, \citenamefont {Colombe}, \citenamefont {Amini}, \citenamefont
  {Leibfried},\ and\ \citenamefont {Wineland}}]{VanDevender2010}%
  \BibitemOpen
  \bibfield  {author} {\bibinfo {author} {\bibfnamefont {A.~P.}\ \bibnamefont
  {VanDevender}}, \bibinfo {author} {\bibfnamefont {Y.}~\bibnamefont
  {Colombe}}, \bibinfo {author} {\bibfnamefont {J.}~\bibnamefont {Amini}},
  \bibinfo {author} {\bibfnamefont {D.}~\bibnamefont {Leibfried}}, \ and\
  \bibinfo {author} {\bibfnamefont {D.~J.}\ \bibnamefont {Wineland}},\
  }\href@noop {} {\bibfield  {journal} {\bibinfo  {journal} {Physical Review
  Letters}\ }\textbf {\bibinfo {volume} {105}},\ \bibinfo {pages} {023001}
  (\bibinfo {year} {2010})}\BibitemShut {NoStop}%
\bibitem [{\citenamefont {Herskind}\ \emph {et~al.}(2010)\citenamefont
  {Herskind}, \citenamefont {Wang}, \citenamefont {Shi}, \citenamefont {Ge},
  \citenamefont {Cetina},\ and\ \citenamefont {Chuang}}]{Herskind2010}%
  \BibitemOpen
  \bibfield  {author} {\bibinfo {author} {\bibfnamefont {P.~F.}\ \bibnamefont
  {Herskind}}, \bibinfo {author} {\bibfnamefont {S.~X.}\ \bibnamefont {Wang}},
  \bibinfo {author} {\bibfnamefont {M.}~\bibnamefont {Shi}}, \bibinfo {author}
  {\bibfnamefont {Y.}~\bibnamefont {Ge}}, \bibinfo {author} {\bibfnamefont
  {M.}~\bibnamefont {Cetina}}, \ and\ \bibinfo {author} {\bibfnamefont {I.~L.}\
  \bibnamefont {Chuang}},\ }\href@noop {} {\bibfield  {journal} {\bibinfo
  {journal} {arXiv:1011.5259v1}\ } (\bibinfo {year} {2010})}\BibitemShut
  {NoStop}%
\bibitem [{\citenamefont {Kim}\ and\ \citenamefont {Kim}(2009)}]{Kim2009}%
  \BibitemOpen
  \bibfield  {author} {\bibinfo {author} {\bibfnamefont {J.}~\bibnamefont
  {Kim}}\ and\ \bibinfo {author} {\bibfnamefont {C.}~\bibnamefont {Kim}},\
  }\href@noop {} {\bibfield  {journal} {\bibinfo  {journal} {Quantum
  Information \& Computation}\ }\textbf {\bibinfo {volume} {9}},\ \bibinfo
  {pages} {0181} (\bibinfo {year} {2009})}\BibitemShut {NoStop}%
\bibitem [{\citenamefont {Schneider}\ \emph {et~al.}(2010)\citenamefont
  {Schneider}, \citenamefont {Enderlein}, \citenamefont {Huber},\ and\
  \citenamefont {Schaetz}}]{SchneiderCh2010}%
  \BibitemOpen
  \bibfield  {author} {\bibinfo {author} {\bibfnamefont {C.}~\bibnamefont
  {Schneider}}, \bibinfo {author} {\bibfnamefont {M.}~\bibnamefont
  {Enderlein}}, \bibinfo {author} {\bibfnamefont {T.}~\bibnamefont {Huber}}, \
  and\ \bibinfo {author} {\bibfnamefont {T.}~\bibnamefont {Schaetz}},\
  }\href@noop {} {\bibfield  {journal} {\bibinfo  {journal} {Nature Photonics}\
  }\textbf {\bibinfo {volume} {4}},\ \bibinfo {pages} {772} (\bibinfo {year}
  {2010})}\BibitemShut {NoStop}%
\bibitem [{\citenamefont {Herskind}\ \emph
  {et~al.}(2009{\natexlab{b}})\citenamefont {Herskind}, \citenamefont {Dantan},
  \citenamefont {Albert}, \citenamefont {Marler},\ and\ \citenamefont
  {Drewsen}}]{Herskind2009}%
  \BibitemOpen
  \bibfield  {author} {\bibinfo {author} {\bibfnamefont {P.~F.}\ \bibnamefont
  {Herskind}}, \bibinfo {author} {\bibfnamefont {A.}~\bibnamefont {Dantan}},
  \bibinfo {author} {\bibfnamefont {M.}~\bibnamefont {Albert}}, \bibinfo
  {author} {\bibfnamefont {J.~P.}\ \bibnamefont {Marler}}, \ and\ \bibinfo
  {author} {\bibfnamefont {M.}~\bibnamefont {Drewsen}},\ }\href
  {http://iopscience.iop.org/0953-4075/42/15/154008/} {\bibfield  {journal}
  {\bibinfo  {journal} {Journal Of Physics B}\ }\textbf {\bibinfo {volume}
  {42}},\ \bibinfo {pages} {154008} (\bibinfo {year}
  {2009}{\natexlab{b}})}\BibitemShut {NoStop}%
\bibitem [{\citenamefont {Blatt}\ and\ \citenamefont
  {Wineland}(2008)}]{Blatt2008}%
  \BibitemOpen
  \bibfield  {author} {\bibinfo {author} {\bibfnamefont {R.}~\bibnamefont
  {Blatt}}\ and\ \bibinfo {author} {\bibfnamefont {D.~J.}\ \bibnamefont
  {Wineland}},\ }\href
  {http://www.nature.com/nature/journal/v453/n7198/abs/nature07125.html}
  {\bibfield  {journal} {\bibinfo  {journal} {Nature}\ }\textbf {\bibinfo
  {volume} {453}},\ \bibinfo {pages} {1008} (\bibinfo {year}
  {2008})}\BibitemShut {NoStop}%
\bibitem [{\citenamefont {Margolis}\ \emph {et~al.}(2004)\citenamefont
  {Margolis}, \citenamefont {Barwood}, \citenamefont {Huang}, \citenamefont
  {Klein}, \citenamefont {Lea}, \citenamefont {Szymaniec},\ and\ \citenamefont
  {Gill}}]{Margolis2004}%
  \BibitemOpen
  \bibfield  {author} {\bibinfo {author} {\bibfnamefont {H.}~\bibnamefont
  {Margolis}}, \bibinfo {author} {\bibfnamefont {G.}~\bibnamefont {Barwood}},
  \bibinfo {author} {\bibfnamefont {G.}~\bibnamefont {Huang}}, \bibinfo
  {author} {\bibfnamefont {H.~A.}\ \bibnamefont {Klein}}, \bibinfo {author}
  {\bibfnamefont {S.}~\bibnamefont {Lea}}, \bibinfo {author} {\bibfnamefont
  {K.}~\bibnamefont {Szymaniec}}, \ and\ \bibinfo {author} {\bibfnamefont
  {P.}~\bibnamefont {Gill}},\ }\href@noop {} {\bibfield  {journal} {\bibinfo
  {journal} {Science}\ }\textbf {\bibinfo {volume} {306}},\ \bibinfo {pages}
  {1355} (\bibinfo {year} {2004})}\BibitemShut {NoStop}%
\bibitem [{\citenamefont {Kim}\ \emph {et~al.}(2010)\citenamefont {Kim},
  \citenamefont {Herskind}, \citenamefont {Kim}, \citenamefont {Kim},\ and\
  \citenamefont {Chuang}}]{KimH2010}%
  \BibitemOpen
  \bibfield  {author} {\bibinfo {author} {\bibfnamefont {T.~H.}\ \bibnamefont
  {Kim}}, \bibinfo {author} {\bibfnamefont {P.~F.}\ \bibnamefont {Herskind}},
  \bibinfo {author} {\bibfnamefont {T.}~\bibnamefont {Kim}}, \bibinfo {author}
  {\bibfnamefont {J.}~\bibnamefont {Kim}}, \ and\ \bibinfo {author}
  {\bibfnamefont {I.~L.}\ \bibnamefont {Chuang}},\ }\href@noop {} {\bibfield
  {journal} {\bibinfo  {journal} {Phys. Rev. A}\ }\textbf {\bibinfo {volume}
  {82}},\ \bibinfo {pages} {043412} (\bibinfo {year} {2010})}\BibitemShut
  {NoStop}%
\bibitem [{\citenamefont {Antohi}\ \emph {et~al.}(2009)\citenamefont {Antohi},
  \citenamefont {Schuster}, \citenamefont {Akselrod}, \citenamefont
  {Labaziewicz}, \citenamefont {Ge}, \citenamefont {Lin}, \citenamefont
  {Bakr},\ and\ \citenamefont {Chuang}}]{Antohi2009}%
  \BibitemOpen
  \bibfield  {author} {\bibinfo {author} {\bibfnamefont {P.~B.}\ \bibnamefont
  {Antohi}}, \bibinfo {author} {\bibfnamefont {D.}~\bibnamefont {Schuster}},
  \bibinfo {author} {\bibfnamefont {G.~M.}\ \bibnamefont {Akselrod}}, \bibinfo
  {author} {\bibfnamefont {J.}~\bibnamefont {Labaziewicz}}, \bibinfo {author}
  {\bibfnamefont {Y.}~\bibnamefont {Ge}}, \bibinfo {author} {\bibfnamefont
  {Z.}~\bibnamefont {Lin}}, \bibinfo {author} {\bibfnamefont {W.~S.}\
  \bibnamefont {Bakr}}, \ and\ \bibinfo {author} {\bibfnamefont {I.~L.}\
  \bibnamefont {Chuang}},\ }\href {http://rsi.aip.org/rsinak/v80/i1/p013103_s1}
  {\bibfield  {journal} {\bibinfo  {journal} {Rev. Sci. Instrum.}\ }\textbf
  {\bibinfo {volume} {80}},\ \bibinfo {pages} {013103} (\bibinfo {year}
  {2009})}\BibitemShut {NoStop}%
\bibitem [{\citenamefont {Brownnutt}\ \emph {et~al.}(2007)\citenamefont
  {Brownnutt}, \citenamefont {Letchumanan}, \citenamefont {Wilpers},
  \citenamefont {Thompson}, \citenamefont {Gill},\ and\ \citenamefont
  {Sinclair}}]{Brownnutt2007}%
  \BibitemOpen
  \bibfield  {author} {\bibinfo {author} {\bibfnamefont {M.}~\bibnamefont
  {Brownnutt}}, \bibinfo {author} {\bibfnamefont {V.}~\bibnamefont
  {Letchumanan}}, \bibinfo {author} {\bibfnamefont {G.}~\bibnamefont
  {Wilpers}}, \bibinfo {author} {\bibfnamefont {R.~C.}\ \bibnamefont
  {Thompson}}, \bibinfo {author} {\bibfnamefont {P.}~\bibnamefont {Gill}}, \
  and\ \bibinfo {author} {\bibfnamefont {A.~G.}\ \bibnamefont {Sinclair}},\
  }\href {http://www.springerlink.com/content/377526u074jh0641/} {\bibfield
  {journal} {\bibinfo  {journal} {Applied Physics B-Lasers And Optics}\
  }\textbf {\bibinfo {volume} {87}},\ \bibinfo {pages} {411} (\bibinfo {year}
  {2007})}\BibitemShut {NoStop}%
\bibitem [{\citenamefont {Dehmelt}(1975)}]{Dehmelt1975}%
  \BibitemOpen
  \bibfield  {author} {\bibinfo {author} {\bibfnamefont {H.}~\bibnamefont
  {Dehmelt}},\ }\href@noop {} {\bibfield  {journal} {\bibinfo  {journal}
  {Bulletin of the American Physical Society}\ }\textbf {\bibinfo {volume}
  {20}},\ \bibinfo {pages} {60} (\bibinfo {year} {1975})}\BibitemShut {NoStop}%
\bibitem [{\citenamefont {Berkeland}\ \emph {et~al.}(1998)\citenamefont
  {Berkeland}, \citenamefont {Miller}, \citenamefont {Bergquist}, \citenamefont
  {Itano},\ and\ \citenamefont {Wineland}}]{Berkeland1998}%
  \BibitemOpen
  \bibfield  {author} {\bibinfo {author} {\bibfnamefont {D.~J.}\ \bibnamefont
  {Berkeland}}, \bibinfo {author} {\bibfnamefont {J.~D.}\ \bibnamefont
  {Miller}}, \bibinfo {author} {\bibfnamefont {J.~C.}\ \bibnamefont
  {Bergquist}}, \bibinfo {author} {\bibfnamefont {W.~M.}\ \bibnamefont
  {Itano}}, \ and\ \bibinfo {author} {\bibfnamefont {D.~J.}\ \bibnamefont
  {Wineland}},\ }\href
  {http://web.ebscohost.com/ehost/detail?vid=1&hid=13&sid=5044e7d6-b010-4e74-8%
ce0-f132b305ba13%40sessionmgr3} {\bibfield  {journal} {\bibinfo  {journal}
  {Journal Of Applied Physics}\ }\textbf {\bibinfo {volume} {83}},\ \bibinfo
  {pages} {5025} (\bibinfo {year} {1998})}\BibitemShut {NoStop}%
\end{thebibliography}


\end{document}